\documentclass[prl,12pt,reprint]{revtex4-1}
\usepackage{amsmath}
\usepackage{graphics}
\usepackage{upgreek}
\usepackage{float}
\usepackage[pdftex]{graphicx}
\usepackage{verbatim}
\newcommand*{\wn}{cm$^{-1}$}
\newcommand*{\Tm}{T$_{2}$}
\newcommand*{\Hm}{H$_{2}$}
\newcommand*{\Dm}{D$_{2}$}

\def\jms{J. Mol.\ Spectrosc.\ }

\begin{document}

\title{Relativistic and QED effects in the fundamental vibration of T$_2$}
\author{T. Madhu Trivikram$^1$, M. Schl\"{o}sser$^2$, W. Ubachs$^1$, and  E. J. Salumbides$^{1}$}
\email{e.j.salumbides@vu.nl}
\affiliation{$^1$Department of Physics and Astronomy, LaserLaB, Vrije Universiteit Amsterdam, De Boelelaan 1081, 1081 HV Amsterdam, The Netherlands\\
$^2$Tritium Laboratory Karlsruhe, Institute of Technical Physics, Karlsruhe Institute of Technology, Hermann-von-Helmholtz-Platz 1, 76344 Eggenstein-Leopoldshafen, Germany}

\date{\today}

\begin{abstract}

The hydrogen molecule has become a test ground for quantum electrodynamical calculations in molecules.
Expanding beyond studies on stable hydrogenic species to the heavier radioactive tritium-bearing molecules, we report on a measurement of the fundamental T$_2$ vibrational splitting $(v= 0 \rightarrow 1)$ for $J=0-5$ rotational levels.
Precision frequency metrology is performed with high-resolution coherent anti-Stokes Raman spectroscopy at an experimental uncertainty of $10-12$~MHz, where sub-Doppler saturation features are exploited for the strongest transition.
The achieved accuracy corresponds to a fifty-fold improvement over a previous measurement, and allows for the extraction of relativistic and QED contributions to T$_2$ transition energies.

\end{abstract}

\maketitle
	
Molecular hydrogen is a quintessential system in the development of quantum chemistry and has emerged as a benchmark for testing relativistic quantum electrodynamics (QED) in simple bound systems.
The accurate measurement of the dissociation energy of the H$_2$  molecule \cite{Liu2009}, the measurement of its fundamental vibrational splitting \cite{Dickenson2013}, as well as the accurate frequency calibration of very weak quadrupole overtone transitions \cite{Campargue2012,Kassi2014,Cheng2012} have been accompanied by ever increasing first principles calculations \cite{Piszczatowski2009,Komasa2011,Pachucki2016}.
The comparisons between accurate theoretical and experimental values have spurred interpretations in fundamental physics, such as contributions from hypothetical fifth forces in the binding of the molecule \cite{Salumbides2013} as well as constraining the compactification lengths of extra dimensions \cite{Salumbides2015b}.
The various contributions to the binding energies in the hydrogen molecule, in particular the adiabatic and nonadiabatic corrections~\cite{Pachucki2015} to the Born-Oppenheimer energies, and to a more subtle extent the relativistic and QED contributions \cite{Puchalski2017}, depend on the masses of the nuclei.
The mass-dependency of the corrections are accentuated in the lightest hydrogenic molecular system, and spectroscopic precision tests were extended to other hydrogen isotopologues.
Measurements of the dissociation energy  \cite{Liu2010} and the quadrupole infrared spectrum \cite{Maddaloni2010,Kassi2012} were extended to the D$_2$ isotopologue, while the mixed HD stable isotopomer was targeted in studies of the dissociation limit \cite{Sprecher2010} and the near infrared spectrum \cite{Kassi2011}.
For performing comparisons with QED calculations the latter were performed for HD as well to high accuracy \cite{Pachucki2010b}.

\begin{figure}[b]
\begin{center}
\includegraphics[width=0.85\linewidth]{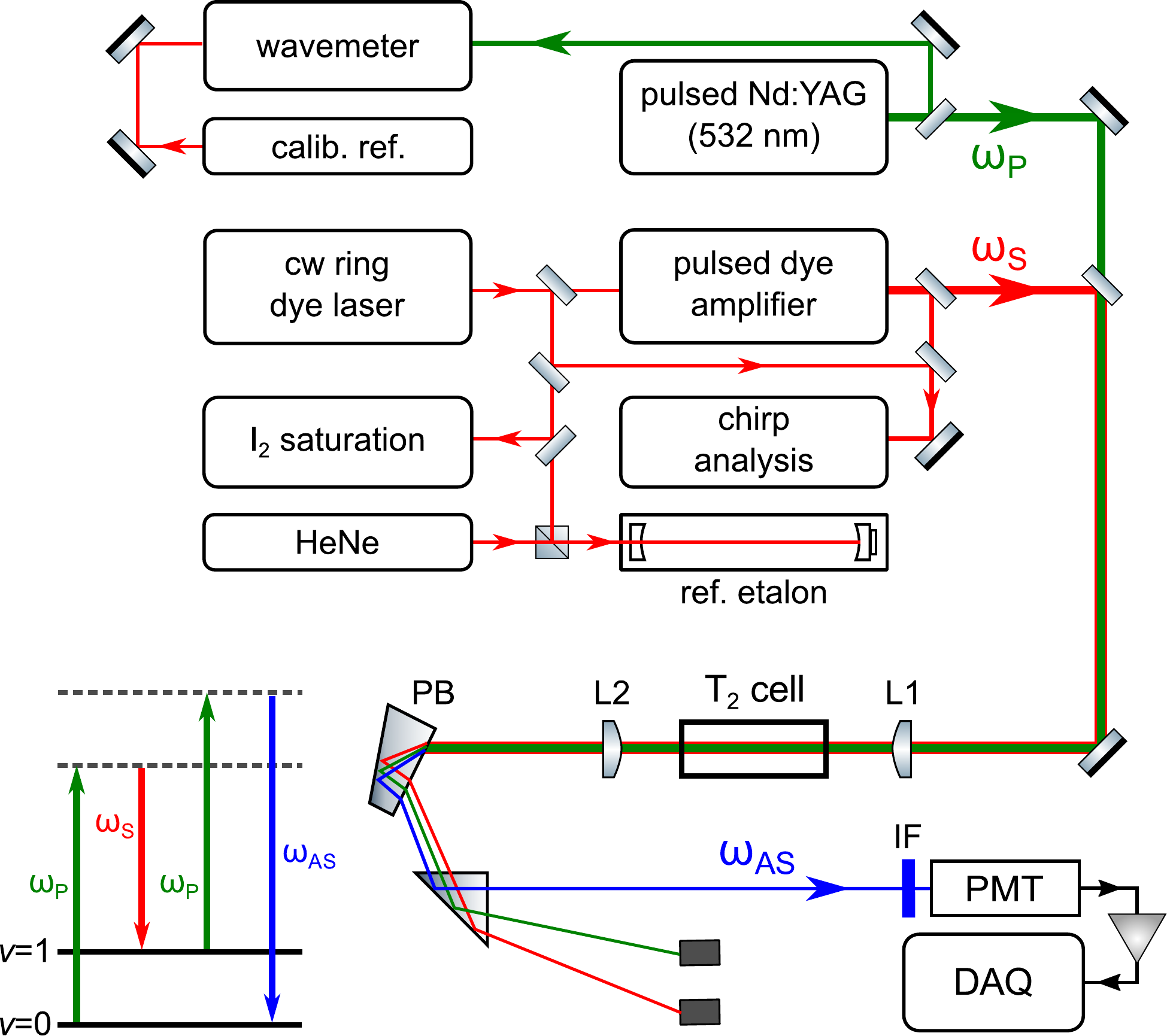}
\caption{\label{setup}
(Color online.) Schematic of the high-resolution CARS setup.
Narrowband pump ($\omega_{P}$) and Stokes ($\omega_{S}$) laser beams are collinearly aligned in a sample cell containing 2.5 mbar of \Tm.
The generated anti-Stokes ($\omega_{AS}$) radiation is spatially dispersed using prisms, passed through an optical interference filter (IF), detected using a photomultiplier tube (PMT) and recorded (DAQ).
A diagram for the CARS frequency-mixing process in shown on the lower left.
}
\end{center}
\end{figure}

In contrast, there is a paucity of high-accuracy investigations on the radioactive tritium-bearing species of molecular hydrogen, such that relativistic and QED effects are entirely untested for the tritiated isotopologues.
Tritium, containing two neutrons in addition to the charge-carrying proton, is unstable with a half-life of about 12 years and undergoes beta decay as the nucleus transmutes from $^{3}$H to $^3$He.
Handling tritium in a typical spectroscopy laboratory is heavily restricted to dilute amounts, thus ruling out the use of molecular beam techniques, while cavity-enhanced techniques face severe difficulty in material degradation with tritium exposure.
Examples of the few gas-phase experiments on T-bearing hydrogen molecules include spontaneous Raman spectroscopy on \Tm~\cite{Edwards1978,Veirs1987} and optoacoustic spectroscopy of the fundamental and overtone bands in HT~\cite{Chuang1987}, performed with sample pressures of a few hundred mbars.
Here, we perform precision tests on \Tm, the heaviest molecular hydrogen species, by employing Coherent Anti-Stokes Raman Spectroscopy (CARS).
CARS offers excellent sensitivity and has been previously applied to \Hm\ at $100$ mbar~\cite{Lucht1989}, while a related Raman technique has been applied to \Dm\ at $2$-mbar pressures~\cite{Owyoung1980}.
We have recently demonstrated the feasibility of precision measurements in a gas cell containing \Tm\ at $2.5$ mbars~\cite{Schloesser2017}.
In this letter, we present results with a fifty-fold increase in precision, obtained by the use of a narrowband Stokes laser source and improved absolute frequency calibrations.
The application of ns-pulsed narrowband laser sources on the low-pressure \Tm\ sample has enabled the observation of narrow sub-Doppler saturation features in the CARS spectra, which is exploited to obtain higher precision for the strongest transition.
Significant enhancement in the detection efficiency also allowed for the use of much lower laser intensities, leading to a more accurate treatment of AC-Stark effects.

\begin{figure}
\begin{center}
\includegraphics[height=0.7\linewidth,width=\linewidth]{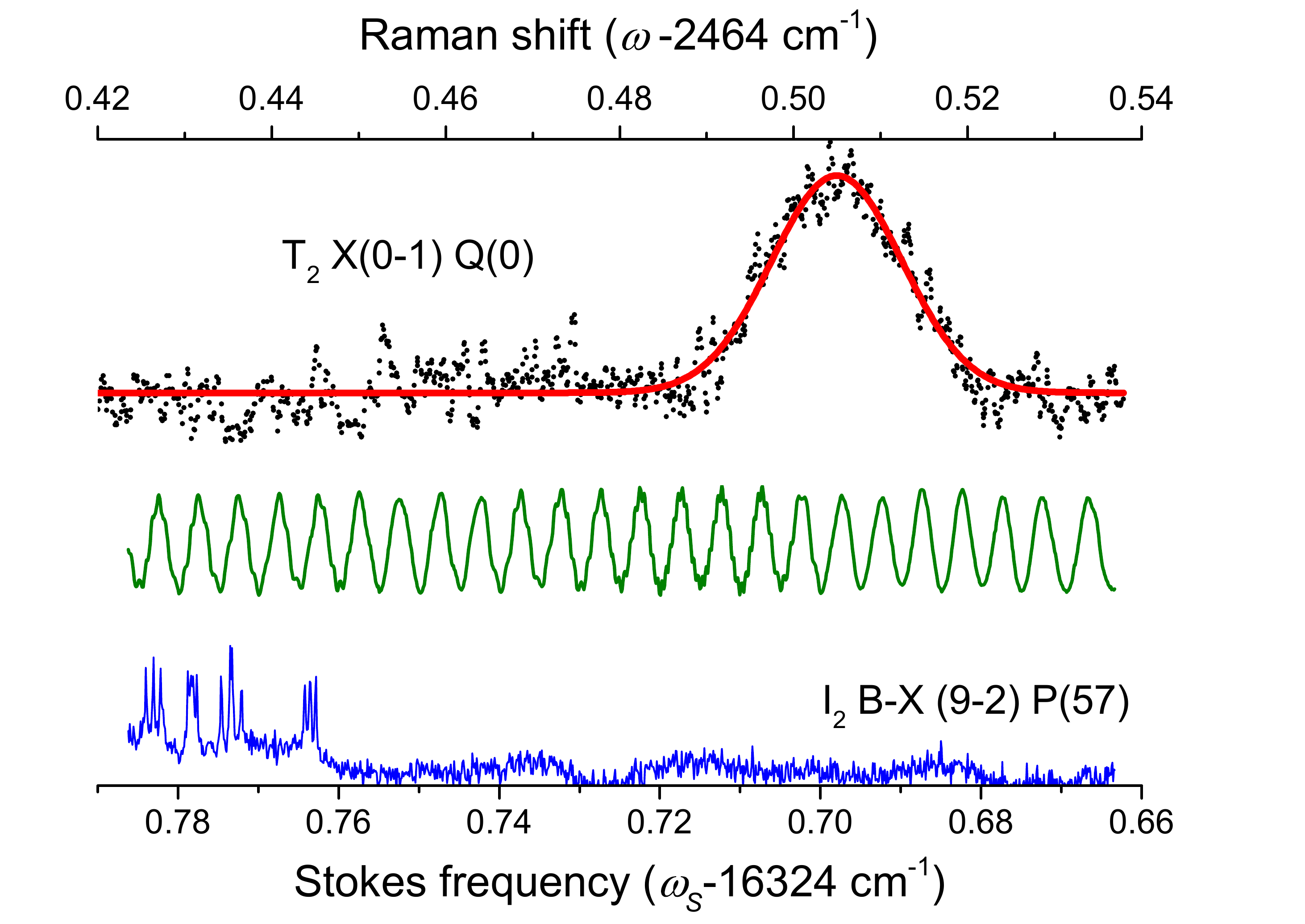}
\caption{\label{specQ0}
(Color online.) Spectral recording of the Q(0) line of the X$^1\Sigma_g^+ (0-1)$ Raman band, probed at peak intensities of 9 GW/cm$^2$ for both lasers, and plotted with respect to the Stokes frequency $\omega_S$ (lower frequency axis) and the Raman shift $\omega$ (upper axis).
The solid red line through the \Tm\ data points is a Gaussian fit, while the transmission of the stabilized etalon and saturated I$_2$ spectrum are plotted below the spectrum for the relative and absolute frequency calibrations for $\omega_S$, respectively.
}
\end{center}
\end{figure}

A schematic representation of the experimental setup is shown in Fig.~\ref{setup}.
Two nearly Fourier-transform limited laser pulses for the pump ($\omega_{P}$, $\lambda =532$ nm) and Stokes ($\omega_{S}$, $\lambda=612$ nm) beams are temporally and spatially overlapped and focused with a $f=20$-cm lens (L1) in the tritium gas cell.
The nonlinear frequency mixing (scheme represented in the lower left corner of Fig.~\ref{setup}) produces an anti-Stokes coherent beam at $\omega_{AS} = 2\omega_{P} - \omega_{S}$ corresponding to $\lambda = 470$ nm, which is collimated (L2: $f=10$ cm) and dispersed using prisms, passed through an optical filter (IF) and finally detected using a photomultiplier tube (PMT).
The pump beam is the output of an injection-seeded and frequency-doubled Nd:YAG laser, while the Stokes radiation is produced using a narrowband pulsed dye amplifier (PDA) system~\cite{Eikema1997}, which is seeded by a continuous-wave (cw) ring dye laser and pumped by a different injection-seeded Nd:YAG laser.
The 4-cm$^3$ gas cell contains 2.5 mbar of mixed molecular hydrogen isotopologues with 93\% \Tm, prepared at the Tritium Laboratory Karlsruhe and transported to LaserLaB Amsterdam~\cite{Schloesser2017}.

\begin{figure}
\begin{center}
\includegraphics[height=0.7\linewidth, width=\linewidth]{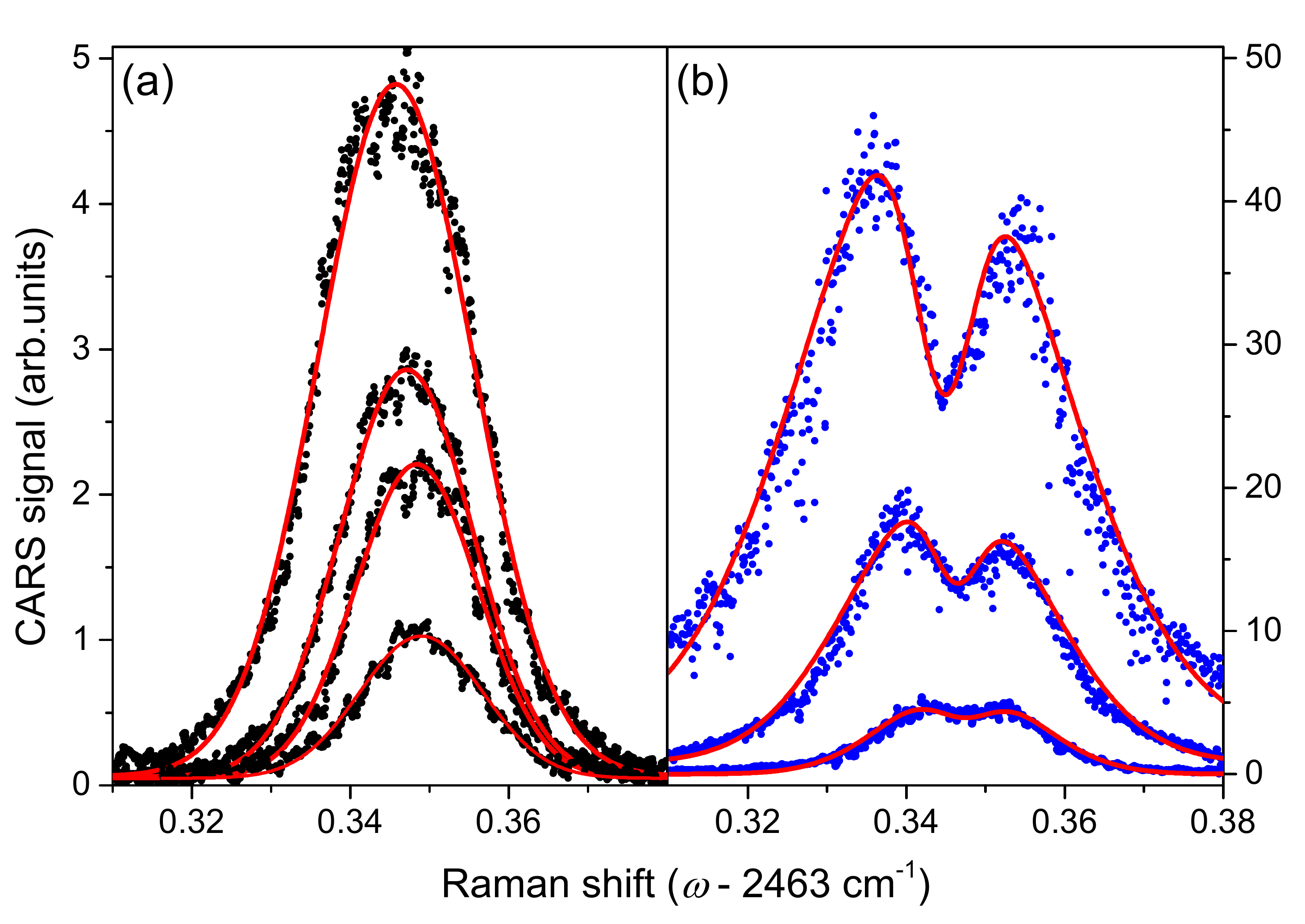}
\caption{\label{peak_shift_sat}
(Color online.) Recorded CARS spectra of the Q(1) line profiles at various $\omega_{P}$ and $\omega_{S}$ laser intensities illustrating the AC-Stark effect.
(a) At low intensities, symmetric profiles are observed that approach towards the instrument- and Doppler-limited bandwidth, with the lowest intensity in (a) at $(I_{P}+I_{S})=6$ GW/cm$^2$.
(b) At sufficiently high intensities, sub-Doppler saturation dips are observed, with the highest intensity in (b) at $(I_{P}+I_{S})=125$ GW/cm$^2$.
The amplitude scale is identical for both panels (a) and (b).
}
\end{center}
\end{figure}

High-resolution CARS spectra are recorded for six Q($J$) rotational lines (for $J=0-5$) of the X$^1\Sigma_g^+$ fundamental $(v=0\rightarrow 1)$ vibrational band.
A typical recording of the Q(0) Raman transition, which is the weakest among the detected lines, is shown in Fig.~\ref{specQ0}.
The cw-seed frequency for the $\omega_S$ radiation is calibrated in scanning mode using transmission markers of a HeNe-stabilized etalon (free spectral range $\nu_\mathrm{FSR}=150.33(1)$ MHz) in combination with a reference spectrum provided from saturation I$_2$ spectroscopy \cite{Xu2000}.
A temporal and spatial cw-pulse frequency offset may be induced by intensity-dependent frequency chirp effects in the pulsed-dye amplification~\cite{Eikema1997,Fee1992,Gangopadhyay1994}, which is measured and corrected for in the data analysis~\cite{Hannemann2006}.
The frequency of the $\omega_P$ pulse is monitored online using a high-resolution wavemeter (High Finesse \r{A}ngstrom WSU-30) that is periodically calibrated against several absolute frequency standards in our laboratory, including calibrations against a Cs standard via an optical frequency comb laser.
The Raman shift $\omega$ is deduced from the simultaneous frequency calibration of both incident lasers at frequencies $\omega_P$ and $\omega_S$, respectively.
 
The spectral lines are typically broadened by the AC-Stark effect as shown in Fig.~\ref{peak_shift_sat}, depending on the pulse intensities of the incident pump $I_{P}$ and Stokes $I_{S}$ laser beams, respectively.
The smallest peak in Fig.~\ref{peak_shift_sat}~(a) was recorded with intensities of $I_{P}=6$ and $I_{S}=1.5$ GW/cm$^2$, and has a full-width half-maximum (FWHM) of 420 MHz.
This approaches the expected linewidth limit from the convolution of the Doppler width ($370$~MHz) with the instrumental bandwidths of both laser beams ($75$~MHz).
At sufficiently high pulse intensities ($>30$ GW/cm$^2$), sub-Doppler CARS saturation dips~\cite{Owyoung1980, Lucht1989} are observed as shown in panel (b) of Fig.~\ref{peak_shift_sat}.
These saturation profiles were fitted with the composite function,
\begin{align*}
y_\mathrm{sat}(\omega) &= A_0 + A_\mathrm{Dopp}\exp{ \left\{ - \left( \frac{\omega-\omega_\mathrm{Dopp}}{\Delta\omega_\mathrm{Dopp}} \right)^2 \right\} }\\
	&\quad - A_\mathrm{sub}\exp{ \left\{ - \left( \frac{\omega-\omega_\mathrm{sub}}{\Delta\omega_\mathrm{sub}} \right)^2 \right\} },
\end{align*}
to obtain the line positions $\omega_{(\cdot)}$, linewidths $\Delta\omega_{(\cdot)}$, and amplitudes $A_{(\cdot)}$ indicated by subscripts (Dopp) and (sub) for the Doppler-broadened profile and sub-Doppler features, respectively.
The lowest intensity scan in Fig.~\ref{peak_shift_sat} (b) shows a resolved sub-Doppler dip with a FWHM linewidth that is four times smaller than the Doppler-limited width and approaches the instrument bandwidth.

\begin{figure}
\begin{center}
\includegraphics[height=0.7\linewidth, width=\linewidth]{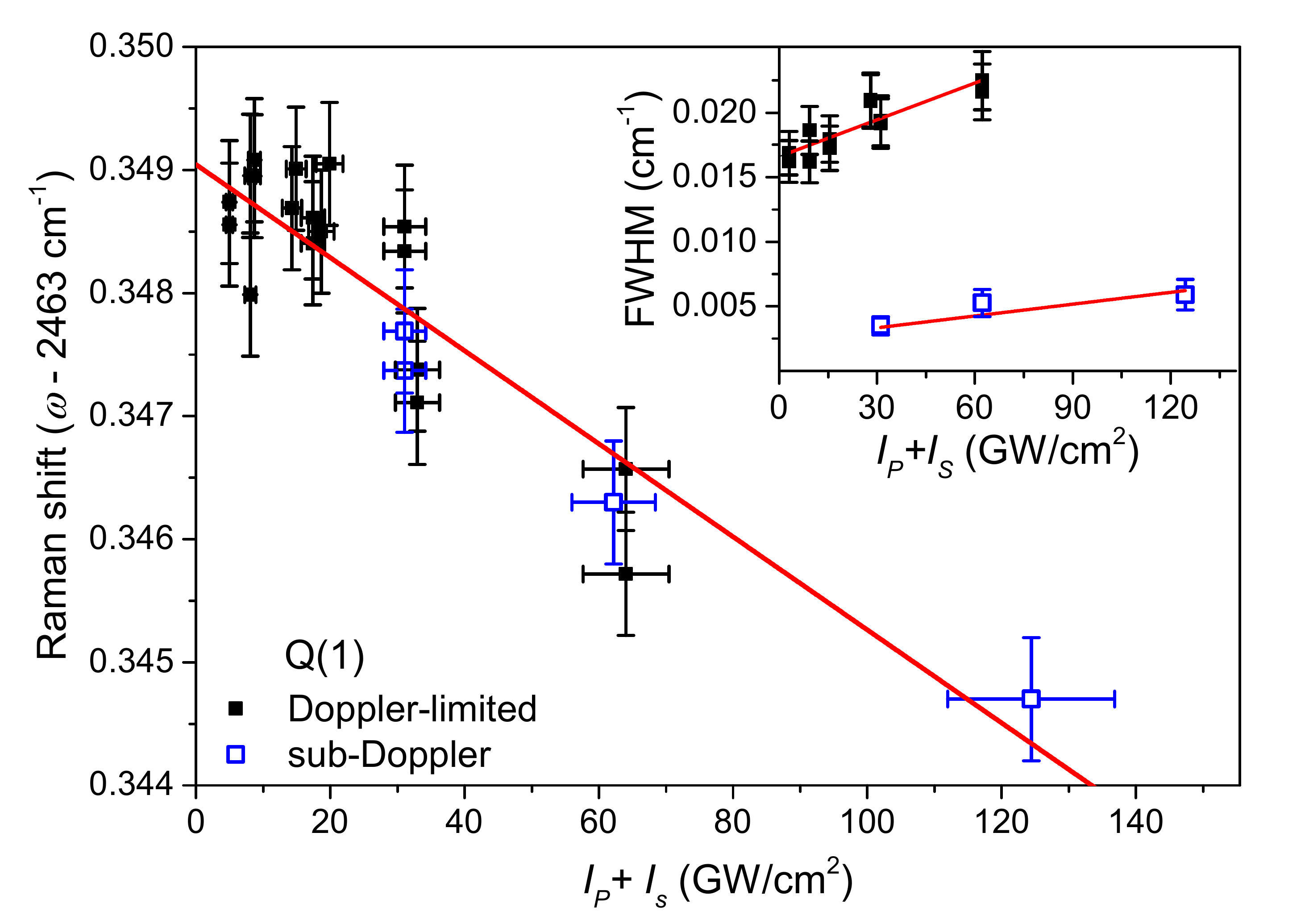}
\caption{\label{stark_Q1}
(Color online.) Extrapolation to the field-free Raman shift for the Q(1) line measured at different intensities.
The solid black squares are obtained from Doppler-broadened spectra while the unfilled blue squares are obtained from the sub-Doppler features in saturated CARS spectra.
The red line is a fit of the combined data sets.
The inset shows the FWHM linewidths, where the sub-Doppler features (solid black squares) are up to four times narrower than those for the Doppler-limited spectra (unfilled blue squares).
}
\end{center}
\end{figure}

The AC-Stark shift for the Q(1) transition is plotted in Fig.~\ref{stark_Q1} as a function of total intensity, $I_{P} + I_{S}$, of both pump and Stokes beams, respectively.
Due to the similar polarizabilities at $\lambda_{P}=532$ nm and $\lambda_{S}=612$ nm for both the $v=0$ and $v=1$ levels for molecular hydrogen~\cite{Dyer1991}, a treatment of the AC-Stark dependence on total intensity was performed.  
The AC-Stark analysis includes the line centers of the Doppler-limited (unsaturated) profiles and sub-Doppler saturation dips, and the true field-free Raman line positions are obtained by extrapolating to zero total intensity.
The Q(1) linewidths are plotted in the inset of Fig.~\ref{stark_Q1} for the Doppler-limited profiles and the saturated sub-Doppler dips, showing the potential of improved line center determinations for the narrow saturation features.
Due to the lower signal to noise ratio for the other Q($J$) lines, sub-Doppler studies were only performed for the Q($1$) transition. 
Collisional shifts in molecular hydrogen have been investigated in CARS studies~\cite{Rahn1990} and are at the level of $\leq 0.1$ MHz at pressures of 2.5 mbar and can be safely ignored for \Tm.
The uncertainty contributions, summarized in Tab.~\ref{tab:unc}, lead to a final uncertainty of $12$ MHz or $4\times10^{-4}$~\wn for Q($J=0,2-5$) lines.
The slightly smaller uncertainty of $10$ MHz for Q($1$) reflects the use of sub-Doppler features in the AC Stark analysis, and better statistics due to more measurements performed on this line for the systematic shift assessment.  
The \emph{statistics} entry in Tab.~\ref{tab:unc} indicates the reproducibility of measurements performed on different days.

\begin{table}[b]
\caption{\label{tab:unc}
Systematic and statistical contributions to the frequency uncertainties in the fundamental vibrational Raman shifts in T$_2$.
Values are given in MHz.}
\begin{ruledtabular}
\begin{tabular}{l@{\hspace{16pt}}r@{\hspace{16pt}}r}
Contribution & \multicolumn{1}{c}{Q($J\neq 1$)} & \multicolumn{1}{c}{Q($1$)}\\
\hline
Pump ($\omega_P$) calibration			& 6  	& 6	\\
Stokes ($\omega_S$) cw calibration		& 2  	& 2	\\
Stokes cw--pulse chirp correction		& 5  	& 5	\\
AC-Stark analysis 				& 7  	& 4	\\		
Collisional shift				& $<1$	& $<1$	\\	
Statistics 					& 7  	& 5	\\		
\hline                                                 
Combined (1$\sigma$)				&12 	&10	\\
\end{tabular}
\end{ruledtabular}
\end{table}

The Q($J$) transition energies for $J=0-5$ are listed in the second column of Tab.~\ref{tab:RelQED}.
The present results are in agreement with our preliminary study~\cite{Schloesser2017}, but represent a fifty-times improvement in accuracy.
The measurements here are also more than two hundred times more precise than all other previous investigations in \Tm, e.g. in Ref.~\cite{Veirs1987}.
The Q($J$) results of the present study are in agreement with the calculations in Ref.~\cite{Schwartz1987} with partial treatment of relativistic and radiative effects, with a claimed accuracy of $0.02$~\wn.
Calculations of the rotationless vibrational splitting Q($0$), which includes relativistic corrections~\cite{Stanke2014} and leading-order QED estimates~\cite{Wolniewicz1993} are also consistent with the present determination to within $0.01$~\wn.

\begin{table}
\caption{\label{tab:RelQED}
Fundamental vibrational splittings of the Q($J$) transitions in \Tm\ obtained in this study are listed in the second column.
Listed in the last column are the relativistic and QED energy contributions, $E_\mathrm{rel+QED}$, to the transition energies extracted from this experiment and nonrelativistic energy $E_\mathrm{nonrel}$ calculations in Ref.~\cite{Pachucki2015}, given in the third column.
Values are given in \wn, with uncertainties in between parentheses.
}
\squeezetable
\begin{ruledtabular}
\begin{tabular}{cccc}
line	& this exp	& $E_\mathrm{nonrel}$~\cite{Pachucki2015}	& $E_\mathrm{rel+QED}$ \\
\hline
Q$(0)$	& 2\,464.5052\,(4) & 2\,464.5021& -0.0031\,(4)\\
Q$(1)$	& 2\,463.3494\,(3) & 2\,463.3463& -0.0031\,(3)\\
Q$(2)$	& 2\,461.0388\,(4) & 2\,461.0372& -0.0016\,(4)\\
Q$(3)$	& 2\,457.5803\,(4) & 2\,457.5795& -0.0008\,(4)\\
Q$(4)$	& 2\,452.9817\,(4) & 2\,452.9803& -0.0014\,(4)\\
Q$(5)$	& 2\,447.2510\,(4) & 2\,447.2492& -0.0017\,(4)\\
\end{tabular}
\end{ruledtabular}
\squeezetable
\end{table}

The nonrelativistic energies, $E_\mathrm{nonrel}$, of the quantum levels in the ground electronic state are now calculated to an accuracy at the level of $10^{-7}$ \wn\ (or kHz-level) for \Hm, \Dm, and \Tm~\cite{Pachucki2015,Pachucki2016}. 
Current efforts in first-principle calculations target higher-order relativistic and QED contributions $E_\mathrm{rel+QED}$, including recoil corrections~\cite{Piszczatowski2009,Komasa2011,Pachucki2016}, and have recently been extended to the $m\alpha^6$-order~\cite{Puchalski2016}.
The evaluation of mass-dependent relativistic nuclear recoil corrections currently dominate the systematic uncertainty of the \emph{ab initio} energies~\cite{Puchalski2017}.
However, the $E_\mathrm{rel+QED}$ contributions to the level energies of \Tm\ have not been calculated to date.
Our measurement accuracy allows for the extraction of $E_\mathrm{rel+QED}$ contributions for \Tm, given in Tab.~\ref{tab:RelQED}, with the use of nonrelativistic level energies $E_\mathrm{nonrel}$ from Ref.~\cite{Pachucki2015}, which may be considered exact for this derivation.

\begin{figure}
\begin{center}
\includegraphics[height=0.7\linewidth, width=\linewidth]{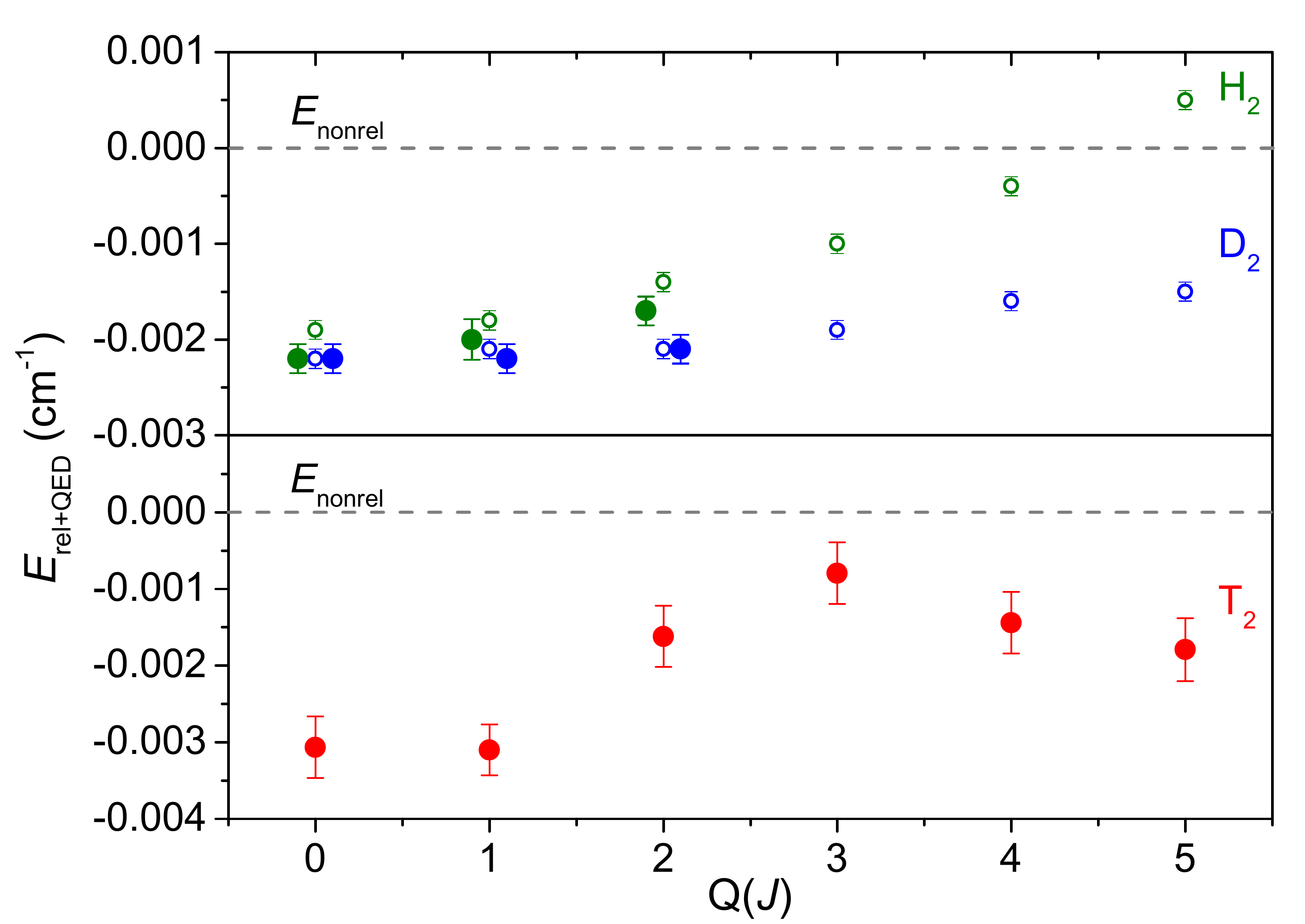}
\caption{\label{RelQED}
(Color online.) Relativistic and QED contributions to \Tm, ($v=0\rightarrow 1$), Q($J$) transition energies, extracted from the present measurements and nonrelativistic calculations in Ref.~\cite{Pachucki2015}, are shown in the lower panel.
The analogous experimentally-derived $E_\mathrm{rel+QED}$ contributions for \Hm\ and \Dm\ from Ref.~\cite{Dickenson2013, Niu2014} are shown in the upper panel, indicated by solid circles and slightly shifted horizontally for clarity.
For \Hm\ and \Dm\ these can be compared to the full \emph{ab initio} calculations~\cite{Komasa2011}, indicated by unfilled circles.
}
\end{center}
\end{figure}

The extracted $E_\mathrm{rel+QED}$ contributions for \Tm\ are plotted in the lower panel of Fig.~\ref{RelQED}.
The analogous contributions of \Hm\ and \Dm\ using the Q($J=0-2$) transitions from molecular beam measurements in Refs.~\cite{Dickenson2013,Niu2014} are plotted in the upper panel.
These experimentally-derived \Hm\ and \Dm\ $E_\mathrm{rel+QED}$ contributions can be compared to the direct \emph{ab initio} calculations~\cite{Komasa2011}, but corresponding \emph{ab initio} calculations for \Tm\ are yet to be carried out.
The relativistic and QED contributions to the \Tm\ transitions measured can be much larger than those for \Hm\ and \Dm, presumably due to the suppression of mass-dependent higher-order terms that scale with the inverse of the reduced mass.

In summary, we have determined Q($J=0-5$) transition energies of the fundamental band of \Tm\ with a 50-fold improvement in precision over our preliminary study~\cite{Schloesser2017} and a 250-times accuracy improvement over all other previous investigations.
The extracted relativistic and QED energy contributions for \Tm\ pose a challenge to high-accuracy calculations that has yet to be pursued.
Access to the tritium-bearing isotopologues (\Tm, HT, DT) doubles the number of the benchmark hydrogen molecule specimens, and greatly expands opportunities for fundamental tests.
Studies using the heavier tritiated species may be useful in disentangling correlations between various mass-dependent effects that currently dominate the calculation uncertainty in molecular hydrogen.
Furthermore, comparisons of experimental and theoretical determinations of transition energies in molecular hydrogen can be used to constrain hypothetical fifth forces~\cite{Salumbides2013}, where the heavier \Tm\ may inherently lead to nine times enhanced sensitivity relative to \Hm.

Future progress in the spectroscopy of molecular tritium holds the promise of a determination of the triton charge radius which is poorly known at present~\cite{Angeli2013}.
This would be of great relevance towards the resolution of the proton size puzzle~\cite{Pohl2010,Antognini2013}, where analogous measurements of the deuteron size~\cite{Pohl2016} have been performed to shed light on the issue.
Advancing our understanding of QED through spectroscopy of tritium-bearing molecular hydrogen may pave the way towards precision studies of nuclear structure.

\begin{acknowledgments}
We would like to acknowledge Meissa Diouf, Adonis Flores and Cunfeng Cheng for assistance during the measurements, and Rob Kortekaas for technical support.
We thank Tobias Falke, David Hillesheimer, Stefan Welte and J\"urgen Wendel of TLK for the preparation and handling of the tritium cell.
WU thanks the European Research Council for an ERC-Advanced grant (No 670168).
The research leading to these results has received funding from LASERLAB-EUROPE (grant agreement no. 654148, European Union’s Horizon 2020 research and innovation programme).
\end{acknowledgments}


\begin{thebibliography}{40}%
\makeatletter
\providecommand \@ifxundefined [1]{%
 \@ifx{#1\undefined}
}%
\providecommand \@ifnum [1]{%
 \ifnum #1\expandafter \@firstoftwo
 \else \expandafter \@secondoftwo
 \fi
}%
\providecommand \@ifx [1]{%
 \ifx #1\expandafter \@firstoftwo
 \else \expandafter \@secondoftwo
 \fi
}%
\providecommand \natexlab [1]{#1}%
\providecommand \enquote  [1]{``#1''}%
\providecommand \bibnamefont  [1]{#1}%
\providecommand \bibfnamefont [1]{#1}%
\providecommand \citenamefont [1]{#1}%
\providecommand \href@noop [0]{\@secondoftwo}%
\providecommand \href [0]{\begingroup \@sanitize@url \@href}%
\providecommand \@href[1]{\@@startlink{#1}\@@href}%
\providecommand \@@href[1]{\endgroup#1\@@endlink}%
\providecommand \@sanitize@url [0]{\catcode `\\12\catcode `\$12\catcode
  `\&12\catcode `\#12\catcode `\^12\catcode `\_12\catcode `\%12\relax}%
\providecommand \@@startlink[1]{}%
\providecommand \@@endlink[0]{}%
\providecommand \url  [0]{\begingroup\@sanitize@url \@url }%
\providecommand \@url [1]{\endgroup\@href {#1}{\urlprefix }}%
\providecommand \urlprefix  [0]{URL }%
\providecommand \Eprint [0]{\href }%
\providecommand \doibase [0]{http://dx.doi.org/}%
\providecommand \selectlanguage [0]{\@gobble}%
\providecommand \bibinfo  [0]{\@secondoftwo}%
\providecommand \bibfield  [0]{\@secondoftwo}%
\providecommand \translation [1]{[#1]}%
\providecommand \BibitemOpen [0]{}%
\providecommand \bibitemStop [0]{}%
\providecommand \bibitemNoStop [0]{.\EOS\space}%
\providecommand \EOS [0]{\spacefactor3000\relax}%
\providecommand \BibitemShut  [1]{\csname bibitem#1\endcsname}%
\let\auto@bib@innerbib\@empty
\bibitem [{\citenamefont {Liu}\ \emph {et~al.}(2009)\citenamefont {Liu},
  \citenamefont {Salumbides}, \citenamefont {Hollenstein}, \citenamefont
  {Koelemeij}, \citenamefont {Eikema}, \citenamefont {Ubachs},\ and\
  \citenamefont {Merkt}}]{Liu2009}%
  \BibitemOpen
  \bibfield  {author} {\bibinfo {author} {\bibfnamefont {J.}~\bibnamefont
  {Liu}}, \bibinfo {author} {\bibfnamefont {E.~J.}\ \bibnamefont {Salumbides}},
  \bibinfo {author} {\bibfnamefont {U.}~\bibnamefont {Hollenstein}}, \bibinfo
  {author} {\bibfnamefont {J.~C.~J.}\ \bibnamefont {Koelemeij}}, \bibinfo
  {author} {\bibfnamefont {K.~S.~E.}\ \bibnamefont {Eikema}}, \bibinfo {author}
  {\bibfnamefont {W.}~\bibnamefont {Ubachs}}, \ and\ \bibinfo {author}
  {\bibfnamefont {F.}~\bibnamefont {Merkt}},\ }\href@noop {} {\bibfield
  {journal} {\bibinfo  {journal} {\jcp}\ }\textbf {\bibinfo {volume} {130}},\
  \bibinfo {pages} {174306} (\bibinfo {year} {2009})}\BibitemShut {NoStop}%
\bibitem [{\citenamefont {Dickenson}\ \emph {et~al.}(2013)\citenamefont
  {Dickenson}, \citenamefont {Niu}, \citenamefont {Salumbides}, \citenamefont
  {Komasa}, \citenamefont {Eikema}, \citenamefont {Pachucki},\ and\
  \citenamefont {Ubachs}}]{Dickenson2013}%
  \BibitemOpen
  \bibfield  {author} {\bibinfo {author} {\bibfnamefont {G.~D.}\ \bibnamefont
  {Dickenson}}, \bibinfo {author} {\bibfnamefont {M.~L.}\ \bibnamefont {Niu}},
  \bibinfo {author} {\bibfnamefont {E.~J.}\ \bibnamefont {Salumbides}},
  \bibinfo {author} {\bibfnamefont {J.}~\bibnamefont {Komasa}}, \bibinfo
  {author} {\bibfnamefont {K.~S.~E.}\ \bibnamefont {Eikema}}, \bibinfo {author}
  {\bibfnamefont {K.}~\bibnamefont {Pachucki}}, \ and\ \bibinfo {author}
  {\bibfnamefont {W.}~\bibnamefont {Ubachs}},\ }\href@noop {} {\bibfield
  {journal} {\bibinfo  {journal} {\prl}\ }\textbf {\bibinfo {volume} {110}},\
  \bibinfo {pages} {193601} (\bibinfo {year} {2013})}\BibitemShut {NoStop}%
\bibitem [{\citenamefont {Campargue}\ \emph {et~al.}(2012)\citenamefont
  {Campargue}, \citenamefont {Kassi}, \citenamefont {Pachucki},\ and\
  \citenamefont {Komasa}}]{Campargue2012}%
  \BibitemOpen
  \bibfield  {author} {\bibinfo {author} {\bibfnamefont {A.}~\bibnamefont
  {Campargue}}, \bibinfo {author} {\bibfnamefont {S.}~\bibnamefont {Kassi}},
  \bibinfo {author} {\bibfnamefont {K.}~\bibnamefont {Pachucki}}, \ and\
  \bibinfo {author} {\bibfnamefont {J.}~\bibnamefont {Komasa}},\ }\href@noop {}
  {\bibfield  {journal} {\bibinfo  {journal} {Phys. Chem. Chem. Phys.}\
  }\textbf {\bibinfo {volume} {14}},\ \bibinfo {pages} {802} (\bibinfo {year}
  {2012})}\BibitemShut {NoStop}%
\bibitem [{\citenamefont {Kassi}\ and\ \citenamefont
  {Campargue}(2014)}]{Kassi2014}%
  \BibitemOpen
  \bibfield  {author} {\bibinfo {author} {\bibfnamefont {S.}~\bibnamefont
  {Kassi}}\ and\ \bibinfo {author} {\bibfnamefont {A.}~\bibnamefont
  {Campargue}},\ }\href@noop {} {\bibfield  {journal} {\bibinfo  {journal}
  {\jms}\ }\textbf {\bibinfo {volume} {300}},\ \bibinfo {pages} {55 } (\bibinfo
  {year} {2014})}\BibitemShut {NoStop}%
\bibitem [{\citenamefont {Cheng}\ \emph {et~al.}(2012)\citenamefont {Cheng},
  \citenamefont {Sun}, \citenamefont {Pan}, \citenamefont {Wang}, \citenamefont
  {Liu}, \citenamefont {Campargue},\ and\ \citenamefont {Hu}}]{Cheng2012}%
  \BibitemOpen
  \bibfield  {author} {\bibinfo {author} {\bibfnamefont {C.-F.}\ \bibnamefont
  {Cheng}}, \bibinfo {author} {\bibfnamefont {Y.~R.}\ \bibnamefont {Sun}},
  \bibinfo {author} {\bibfnamefont {H.}~\bibnamefont {Pan}}, \bibinfo {author}
  {\bibfnamefont {J.}~\bibnamefont {Wang}}, \bibinfo {author} {\bibfnamefont
  {A.-W.}\ \bibnamefont {Liu}}, \bibinfo {author} {\bibfnamefont
  {A.}~\bibnamefont {Campargue}}, \ and\ \bibinfo {author} {\bibfnamefont
  {S.-M.}\ \bibnamefont {Hu}},\ }\href@noop {} {\bibfield  {journal} {\bibinfo
  {journal} {\pra}\ }\textbf {\bibinfo {volume} {85}},\ \bibinfo {pages}
  {024501} (\bibinfo {year} {2012})}\BibitemShut {NoStop}%
\bibitem [{\citenamefont {Piszczatowski}\ \emph {et~al.}(2009)\citenamefont
  {Piszczatowski}, \citenamefont {\L{}ach}, \citenamefont {Przybytek},
  \citenamefont {Komasa}, \citenamefont {Pachucki},\ and\ \citenamefont
  {Jeziorski}}]{Piszczatowski2009}%
  \BibitemOpen
  \bibfield  {author} {\bibinfo {author} {\bibfnamefont {K.}~\bibnamefont
  {Piszczatowski}}, \bibinfo {author} {\bibfnamefont {G.}~\bibnamefont
  {\L{}ach}}, \bibinfo {author} {\bibfnamefont {M.}~\bibnamefont {Przybytek}},
  \bibinfo {author} {\bibfnamefont {J.}~\bibnamefont {Komasa}}, \bibinfo
  {author} {\bibfnamefont {K.}~\bibnamefont {Pachucki}}, \ and\ \bibinfo
  {author} {\bibfnamefont {B.}~\bibnamefont {Jeziorski}},\ }\href@noop {}
  {\bibfield  {journal} {\bibinfo  {journal} {J. Chem. Theory Comput.}\
  }\textbf {\bibinfo {volume} {5}},\ \bibinfo {pages} {3039} (\bibinfo {year}
  {2009})}\BibitemShut {NoStop}%
\bibitem [{\citenamefont {Komasa}\ \emph {et~al.}(2011)\citenamefont {Komasa},
  \citenamefont {Piszczatowski}, \citenamefont {\L{}ach}, \citenamefont
  {Przybytek}, \citenamefont {Jeziorski},\ and\ \citenamefont
  {Pachucki}}]{Komasa2011}%
  \BibitemOpen
  \bibfield  {author} {\bibinfo {author} {\bibfnamefont {J.}~\bibnamefont
  {Komasa}}, \bibinfo {author} {\bibfnamefont {K.}~\bibnamefont
  {Piszczatowski}}, \bibinfo {author} {\bibfnamefont {G.}~\bibnamefont
  {\L{}ach}}, \bibinfo {author} {\bibfnamefont {M.}~\bibnamefont {Przybytek}},
  \bibinfo {author} {\bibfnamefont {B.}~\bibnamefont {Jeziorski}}, \ and\
  \bibinfo {author} {\bibfnamefont {K.}~\bibnamefont {Pachucki}},\ }\href@noop
  {} {\bibfield  {journal} {\bibinfo  {journal} {J. Chem. Theory Comput.}\
  }\textbf {\bibinfo {volume} {7}},\ \bibinfo {pages} {3105} (\bibinfo {year}
  {2011})}\BibitemShut {NoStop}%
\bibitem [{\citenamefont {Pachucki}\ and\ \citenamefont
  {Komasa}(2016)}]{Pachucki2016}%
  \BibitemOpen
  \bibfield  {author} {\bibinfo {author} {\bibfnamefont {K.}~\bibnamefont
  {Pachucki}}\ and\ \bibinfo {author} {\bibfnamefont {J.}~\bibnamefont
  {Komasa}},\ }\href@noop {} {\bibfield  {journal} {\bibinfo  {journal} {J.
  Chem. Phys.}\ }\textbf {\bibinfo {volume} {144}},\ \bibinfo {pages} {164306}
  (\bibinfo {year} {2016})}\BibitemShut {NoStop}%
\bibitem [{\citenamefont {Salumbides}\ \emph {et~al.}(2013)\citenamefont
  {Salumbides}, \citenamefont {Koelemeij}, \citenamefont {Komasa},
  \citenamefont {Pachucki}, \citenamefont {Eikema},\ and\ \citenamefont
  {Ubachs}}]{Salumbides2013}%
  \BibitemOpen
  \bibfield  {author} {\bibinfo {author} {\bibfnamefont {E.~J.}\ \bibnamefont
  {Salumbides}}, \bibinfo {author} {\bibfnamefont {J.~C.~J.}\ \bibnamefont
  {Koelemeij}}, \bibinfo {author} {\bibfnamefont {J.}~\bibnamefont {Komasa}},
  \bibinfo {author} {\bibfnamefont {K.}~\bibnamefont {Pachucki}}, \bibinfo
  {author} {\bibfnamefont {K.~S.~E.}\ \bibnamefont {Eikema}}, \ and\ \bibinfo
  {author} {\bibfnamefont {W.}~\bibnamefont {Ubachs}},\ }\href@noop {}
  {\bibfield  {journal} {\bibinfo  {journal} {\prd}\ }\textbf {\bibinfo
  {volume} {87}},\ \bibinfo {pages} {112008} (\bibinfo {year}
  {2013})}\BibitemShut {NoStop}%
\bibitem [{\citenamefont {Salumbides}\ \emph {et~al.}(2015)\citenamefont
  {Salumbides}, \citenamefont {Schellekens}, \citenamefont {Gato-Rivera},\ and\
  \citenamefont {Ubachs}}]{Salumbides2015b}%
  \BibitemOpen
  \bibfield  {author} {\bibinfo {author} {\bibfnamefont {E.~J.}\ \bibnamefont
  {Salumbides}}, \bibinfo {author} {\bibfnamefont {A.~N.}\ \bibnamefont
  {Schellekens}}, \bibinfo {author} {\bibfnamefont {B.}~\bibnamefont
  {Gato-Rivera}}, \ and\ \bibinfo {author} {\bibfnamefont {W.}~\bibnamefont
  {Ubachs}},\ }\href@noop {} {\bibfield  {journal} {\bibinfo  {journal} {New J.
  Phys.}\ }\textbf {\bibinfo {volume} {17}},\ \bibinfo {pages} {033015}
  (\bibinfo {year} {2015})}\BibitemShut {NoStop}%
\bibitem [{\citenamefont {Pachucki}\ and\ \citenamefont
  {Komasa}(2015)}]{Pachucki2015}%
  \BibitemOpen
  \bibfield  {author} {\bibinfo {author} {\bibfnamefont {K.}~\bibnamefont
  {Pachucki}}\ and\ \bibinfo {author} {\bibfnamefont {J.}~\bibnamefont
  {Komasa}},\ }\href@noop {} {\bibfield  {journal} {\bibinfo  {journal} {J.
  Chem. Phys.}\ }\textbf {\bibinfo {volume} {143}},\ \bibinfo {pages} {034111}
  (\bibinfo {year} {2015})}\BibitemShut {NoStop}%
\bibitem [{\citenamefont {Puchalski}\ \emph {et~al.}(2017)\citenamefont
  {Puchalski}, \citenamefont {Komasa},\ and\ \citenamefont
  {Pachucki}}]{Puchalski2017}%
  \BibitemOpen
  \bibfield  {author} {\bibinfo {author} {\bibfnamefont {M.}~\bibnamefont
  {Puchalski}}, \bibinfo {author} {\bibfnamefont {J.}~\bibnamefont {Komasa}}, \
  and\ \bibinfo {author} {\bibfnamefont {K.}~\bibnamefont {Pachucki}},\
  }\href@noop {} {\bibfield  {journal} {\bibinfo  {journal} {Phys. Rev. A}\
  }\textbf {\bibinfo {volume} {95}},\ \bibinfo {pages} {052506} (\bibinfo
  {year} {2017})}\BibitemShut {NoStop}%
\bibitem [{\citenamefont {Liu}\ \emph {et~al.}(2010)\citenamefont {Liu},
  \citenamefont {Sprecher}, \citenamefont {Jungen}, \citenamefont {Ubachs},\
  and\ \citenamefont {Merkt}}]{Liu2010}%
  \BibitemOpen
  \bibfield  {author} {\bibinfo {author} {\bibfnamefont {J.}~\bibnamefont
  {Liu}}, \bibinfo {author} {\bibfnamefont {D.}~\bibnamefont {Sprecher}},
  \bibinfo {author} {\bibfnamefont {C.}~\bibnamefont {Jungen}}, \bibinfo
  {author} {\bibfnamefont {W.}~\bibnamefont {Ubachs}}, \ and\ \bibinfo {author}
  {\bibfnamefont {F.}~\bibnamefont {Merkt}},\ }\href@noop {} {\bibfield
  {journal} {\bibinfo  {journal} {\jcp}\ }\textbf {\bibinfo {volume} {132}},\
  \bibinfo {pages} {154301} (\bibinfo {year} {2010})}\BibitemShut {NoStop}%
\bibitem [{\citenamefont {Maddaloni}\ \emph {et~al.}(2010)\citenamefont
  {Maddaloni}, \citenamefont {Malara}, \citenamefont {Tommasi}, \citenamefont
  {Rosa}, \citenamefont {Ricciardi}, \citenamefont {Gagliardi}, \citenamefont
  {Tamassia}, \citenamefont {Lonardo},\ and\ \citenamefont
  {Natale}}]{Maddaloni2010}%
  \BibitemOpen
  \bibfield  {author} {\bibinfo {author} {\bibfnamefont {P.}~\bibnamefont
  {Maddaloni}}, \bibinfo {author} {\bibfnamefont {P.}~\bibnamefont {Malara}},
  \bibinfo {author} {\bibfnamefont {E.~D.}\ \bibnamefont {Tommasi}}, \bibinfo
  {author} {\bibfnamefont {M.~D.}\ \bibnamefont {Rosa}}, \bibinfo {author}
  {\bibfnamefont {I.}~\bibnamefont {Ricciardi}}, \bibinfo {author}
  {\bibfnamefont {G.}~\bibnamefont {Gagliardi}}, \bibinfo {author}
  {\bibfnamefont {F.}~\bibnamefont {Tamassia}}, \bibinfo {author}
  {\bibfnamefont {G.~D.}\ \bibnamefont {Lonardo}}, \ and\ \bibinfo {author}
  {\bibfnamefont {P.~D.}\ \bibnamefont {Natale}},\ }\href@noop {} {\bibfield
  {journal} {\bibinfo  {journal} {J. Chem. Phys.}\ }\textbf {\bibinfo {volume}
  {133}},\ \bibinfo {pages} {154317} (\bibinfo {year} {2010})}\BibitemShut
  {NoStop}%
\bibitem [{\citenamefont {Kassi}\ \emph {et~al.}(2012)\citenamefont {Kassi},
  \citenamefont {Campargue}, \citenamefont {Pachucki},\ and\ \citenamefont
  {Komasa}}]{Kassi2012}%
  \BibitemOpen
  \bibfield  {author} {\bibinfo {author} {\bibfnamefont {S.}~\bibnamefont
  {Kassi}}, \bibinfo {author} {\bibfnamefont {A.}~\bibnamefont {Campargue}},
  \bibinfo {author} {\bibfnamefont {K.}~\bibnamefont {Pachucki}}, \ and\
  \bibinfo {author} {\bibfnamefont {J.}~\bibnamefont {Komasa}},\ }\href@noop {}
  {\bibfield  {journal} {\bibinfo  {journal} {J. Chem. Phys.}\ }\textbf
  {\bibinfo {volume} {136}},\ \bibinfo {pages} {184309} (\bibinfo {year}
  {2012})}\BibitemShut {NoStop}%
\bibitem [{\citenamefont {Sprecher}\ \emph {et~al.}(2010)\citenamefont
  {Sprecher}, \citenamefont {Liu}, \citenamefont {Jungen}, \citenamefont
  {Ubachs},\ and\ \citenamefont {Merkt}}]{Sprecher2010}%
  \BibitemOpen
  \bibfield  {author} {\bibinfo {author} {\bibfnamefont {D.}~\bibnamefont
  {Sprecher}}, \bibinfo {author} {\bibfnamefont {J.}~\bibnamefont {Liu}},
  \bibinfo {author} {\bibfnamefont {C.}~\bibnamefont {Jungen}}, \bibinfo
  {author} {\bibfnamefont {W.}~\bibnamefont {Ubachs}}, \ and\ \bibinfo {author}
  {\bibfnamefont {F.}~\bibnamefont {Merkt}},\ }\href@noop {} {\bibfield
  {journal} {\bibinfo  {journal} {J. Chem. Phys.}\ }\textbf {\bibinfo {volume}
  {133}},\ \bibinfo {pages} {111102} (\bibinfo {year} {2010})}\BibitemShut
  {NoStop}%
\bibitem [{\citenamefont {Kassi}\ and\ \citenamefont
  {Campargue}(2011)}]{Kassi2011}%
  \BibitemOpen
  \bibfield  {author} {\bibinfo {author} {\bibfnamefont {S.}~\bibnamefont
  {Kassi}}\ and\ \bibinfo {author} {\bibfnamefont {A.}~\bibnamefont
  {Campargue}},\ }\href@noop {} {\bibfield  {journal} {\bibinfo  {journal} {J.
  Mol. Spectrosc.}\ }\textbf {\bibinfo {volume} {267}},\ \bibinfo {pages} {36}
  (\bibinfo {year} {2011})}\BibitemShut {NoStop}%
\bibitem [{\citenamefont {Pachucki}\ and\ \citenamefont
  {Komasa}(2010)}]{Pachucki2010b}%
  \BibitemOpen
  \bibfield  {author} {\bibinfo {author} {\bibfnamefont {K.}~\bibnamefont
  {Pachucki}}\ and\ \bibinfo {author} {\bibfnamefont {J.}~\bibnamefont
  {Komasa}},\ }\href@noop {} {\bibfield  {journal} {\bibinfo  {journal} {Phys.
  Chem. Chem. Phys.}\ }\textbf {\bibinfo {volume} {12}},\ \bibinfo {pages}
  {9188} (\bibinfo {year} {2010})}\BibitemShut {NoStop}%
\bibitem [{\citenamefont {Edwards}\ \emph {et~al.}(1978)\citenamefont
  {Edwards}, \citenamefont {Long},\ and\ \citenamefont
  {Mansour}}]{Edwards1978}%
  \BibitemOpen
  \bibfield  {author} {\bibinfo {author} {\bibfnamefont {H.~G.~M.}\
  \bibnamefont {Edwards}}, \bibinfo {author} {\bibfnamefont {D.~A.}\
  \bibnamefont {Long}}, \ and\ \bibinfo {author} {\bibfnamefont {H.~R.}\
  \bibnamefont {Mansour}},\ }\href@noop {} {\bibfield  {journal} {\bibinfo
  {journal} {J. Chem. Soc.{,} Faraday Trans. 2}\ }\textbf {\bibinfo {volume}
  {74}},\ \bibinfo {pages} {1203} (\bibinfo {year} {1978})}\BibitemShut
  {NoStop}%
\bibitem [{\citenamefont {Veirs}\ and\ \citenamefont
  {Rosenblatt}(1987)}]{Veirs1987}%
  \BibitemOpen
  \bibfield  {author} {\bibinfo {author} {\bibfnamefont {D.}~\bibnamefont
  {Veirs}}\ and\ \bibinfo {author} {\bibfnamefont {G.}~\bibnamefont
  {Rosenblatt}},\ }\href@noop {} {\bibfield  {journal} {\bibinfo  {journal} {J.
  Mol. Spectrosc.}\ }\textbf {\bibinfo {volume} {121}},\ \bibinfo {pages} {401}
  (\bibinfo {year} {1987})}\BibitemShut {NoStop}%
\bibitem [{\citenamefont {Chuang}\ and\ \citenamefont
  {Zare}(1987)}]{Chuang1987}%
  \BibitemOpen
  \bibfield  {author} {\bibinfo {author} {\bibfnamefont {M.-C.}\ \bibnamefont
  {Chuang}}\ and\ \bibinfo {author} {\bibfnamefont {R.~N.}\ \bibnamefont
  {Zare}},\ }\href@noop {} {\bibfield  {journal} {\bibinfo  {journal} {J. Mol.
  Spectrosc.}\ }\textbf {\bibinfo {volume} {121}},\ \bibinfo {pages} {380 }
  (\bibinfo {year} {1987})}\BibitemShut {NoStop}%
\bibitem [{\citenamefont {Lucht}\ and\ \citenamefont
  {Farrow}(1989)}]{Lucht1989}%
  \BibitemOpen
  \bibfield  {author} {\bibinfo {author} {\bibfnamefont {R.~P.}\ \bibnamefont
  {Lucht}}\ and\ \bibinfo {author} {\bibfnamefont {R.~L.}\ \bibnamefont
  {Farrow}},\ }\href {\doibase 10.1364/JOSAB.6.002313} {\bibfield  {journal}
  {\bibinfo  {journal} {J. Opt. Soc. Am. B}\ }\textbf {\bibinfo {volume} {6}},\
  \bibinfo {pages} {2313} (\bibinfo {year} {1989})}\BibitemShut {NoStop}%
\bibitem [{\citenamefont {Owyoung}\ and\ \citenamefont
  {Esherick}(1980)}]{Owyoung1980}%
  \BibitemOpen
  \bibfield  {author} {\bibinfo {author} {\bibfnamefont {A.}~\bibnamefont
  {Owyoung}}\ and\ \bibinfo {author} {\bibfnamefont {P.}~\bibnamefont
  {Esherick}},\ }\href@noop {} {\bibfield  {journal} {\bibinfo  {journal} {Opt.
  Lett.}\ }\textbf {\bibinfo {volume} {5}},\ \bibinfo {pages} {421} (\bibinfo
  {year} {1980})}\BibitemShut {NoStop}%
\bibitem [{\citenamefont {Schl\"{o}sser}\ \emph {et~al.}(2017)\citenamefont
  {Schl\"{o}sser}, \citenamefont {Zhao}, \citenamefont {Trivikram},
  \citenamefont {Ubachs},\ and\ \citenamefont {Salumbides}}]{Schloesser2017}%
  \BibitemOpen
  \bibfield  {author} {\bibinfo {author} {\bibfnamefont {M.}~\bibnamefont
  {Schl\"{o}sser}}, \bibinfo {author} {\bibfnamefont {X.}~\bibnamefont {Zhao}},
  \bibinfo {author} {\bibfnamefont {M.~T.}\ \bibnamefont {Trivikram}}, \bibinfo
  {author} {\bibfnamefont {W.}~\bibnamefont {Ubachs}}, \ and\ \bibinfo {author}
  {\bibfnamefont {E.~J.}\ \bibnamefont {Salumbides}},\ }\href@noop {}
  {\bibfield  {journal} {\bibinfo  {journal} {J. Phys. B}\ }\textbf {\bibinfo
  {volume} {50}},\ \bibinfo {pages} {214004} (\bibinfo {year}
  {2017})}\BibitemShut {NoStop}%
\bibitem [{\citenamefont {Eikema}\ \emph {et~al.}(1997)\citenamefont {Eikema},
  \citenamefont {Ubachs}, \citenamefont {Vassen},\ and\ \citenamefont
  {Hogervorst}}]{Eikema1997}%
  \BibitemOpen
  \bibfield  {author} {\bibinfo {author} {\bibfnamefont {K.~S.~E.}\
  \bibnamefont {Eikema}}, \bibinfo {author} {\bibfnamefont {W.}~\bibnamefont
  {Ubachs}}, \bibinfo {author} {\bibfnamefont {W.}~\bibnamefont {Vassen}}, \
  and\ \bibinfo {author} {\bibfnamefont {W.}~\bibnamefont {Hogervorst}},\
  }\href@noop {} {\bibfield  {journal} {\bibinfo  {journal} {Phys. Rev. A}\
  }\textbf {\bibinfo {volume} {55}},\ \bibinfo {pages} {1866} (\bibinfo {year}
  {1997})}\BibitemShut {NoStop}%
\bibitem [{\citenamefont {Xu}\ \emph {et~al.}(2000)\citenamefont {Xu},
  \citenamefont {van Dierendonck}, \citenamefont {Hogervorst},\ and\
  \citenamefont {Ubachs}}]{Xu2000}%
  \BibitemOpen
  \bibfield  {author} {\bibinfo {author} {\bibfnamefont {S.}~\bibnamefont
  {Xu}}, \bibinfo {author} {\bibfnamefont {R.}~\bibnamefont {van Dierendonck}},
  \bibinfo {author} {\bibfnamefont {W.}~\bibnamefont {Hogervorst}}, \ and\
  \bibinfo {author} {\bibfnamefont {W.}~\bibnamefont {Ubachs}},\ }\href@noop {}
  {\bibfield  {journal} {\bibinfo  {journal} {\jms}\ }\textbf {\bibinfo
  {volume} {201}},\ \bibinfo {pages} {256} (\bibinfo {year}
  {2000})}\BibitemShut {NoStop}%
\bibitem [{\citenamefont {Fee}\ \emph {et~al.}(1992)\citenamefont {Fee},
  \citenamefont {Danzmann},\ and\ \citenamefont {Chu}}]{Fee1992}%
  \BibitemOpen
  \bibfield  {author} {\bibinfo {author} {\bibfnamefont {M.~S.}\ \bibnamefont
  {Fee}}, \bibinfo {author} {\bibfnamefont {K.}~\bibnamefont {Danzmann}}, \
  and\ \bibinfo {author} {\bibfnamefont {S.}~\bibnamefont {Chu}},\ }\href@noop
  {} {\bibfield  {journal} {\bibinfo  {journal} {Phys. Rev. A}\ }\textbf
  {\bibinfo {volume} {45}},\ \bibinfo {pages} {4911} (\bibinfo {year}
  {1992})}\BibitemShut {NoStop}%
\bibitem [{\citenamefont {Gangopadhyay}\ \emph {et~al.}(1994)\citenamefont
  {Gangopadhyay}, \citenamefont {Melikechi},\ and\ \citenamefont
  {Eyler}}]{Gangopadhyay1994}%
  \BibitemOpen
  \bibfield  {author} {\bibinfo {author} {\bibfnamefont {S.}~\bibnamefont
  {Gangopadhyay}}, \bibinfo {author} {\bibfnamefont {N.}~\bibnamefont
  {Melikechi}}, \ and\ \bibinfo {author} {\bibfnamefont {E.~E.}\ \bibnamefont
  {Eyler}},\ }\href@noop {} {\bibfield  {journal} {\bibinfo  {journal} {J. Opt.
  Soc. Am. B}\ }\textbf {\bibinfo {volume} {11}},\ \bibinfo {pages} {231}
  (\bibinfo {year} {1994})}\BibitemShut {NoStop}%
\bibitem [{\citenamefont {Hannemann}\ \emph {et~al.}(2006)\citenamefont
  {Hannemann}, \citenamefont {Salumbides}, \citenamefont {Witte}, \citenamefont
  {Zinkstok}, \citenamefont {van Duijn}, \citenamefont {Eikema},\ and\
  \citenamefont {Ubachs}}]{Hannemann2006}%
  \BibitemOpen
  \bibfield  {author} {\bibinfo {author} {\bibfnamefont {S.}~\bibnamefont
  {Hannemann}}, \bibinfo {author} {\bibfnamefont {E.~J.}\ \bibnamefont
  {Salumbides}}, \bibinfo {author} {\bibfnamefont {S.}~\bibnamefont {Witte}},
  \bibinfo {author} {\bibfnamefont {R.~T.}\ \bibnamefont {Zinkstok}}, \bibinfo
  {author} {\bibfnamefont {E.~J.}\ \bibnamefont {van Duijn}}, \bibinfo {author}
  {\bibfnamefont {K.~S.~E.}\ \bibnamefont {Eikema}}, \ and\ \bibinfo {author}
  {\bibfnamefont {W.}~\bibnamefont {Ubachs}},\ }\href@noop {} {\bibfield
  {journal} {\bibinfo  {journal} {Phys. Rev. A}\ }\textbf {\bibinfo {volume}
  {74}},\ \bibinfo {pages} {062514} (\bibinfo {year} {2006})}\BibitemShut
  {NoStop}%
\bibitem [{\citenamefont {Dyer}\ and\ \citenamefont
  {Bischel}(1991)}]{Dyer1991}%
  \BibitemOpen
  \bibfield  {author} {\bibinfo {author} {\bibfnamefont {M.~J.}\ \bibnamefont
  {Dyer}}\ and\ \bibinfo {author} {\bibfnamefont {W.~K.}\ \bibnamefont
  {Bischel}},\ }\href@noop {} {\bibfield  {journal} {\bibinfo  {journal} {Phys.
  Rev. A}\ }\textbf {\bibinfo {volume} {44}},\ \bibinfo {pages} {3138}
  (\bibinfo {year} {1991})}\BibitemShut {NoStop}%
\bibitem [{\citenamefont {Rahn}\ and\ \citenamefont
  {Rosasco}(1990)}]{Rahn1990}%
  \BibitemOpen
  \bibfield  {author} {\bibinfo {author} {\bibfnamefont {L.~A.}~\bibnamefont
  {Rahn}}\ and\ \bibinfo {author} {\bibfnamefont {G.~J.}~\bibnamefont {Rosasco}},\
  }\href@noop {} {\bibfield  {journal} {\bibinfo  {journal} {Phys. Rev. A}\
  }\textbf {\bibinfo {volume} {41}},\ \bibinfo {pages} {3698} (\bibinfo {year}
  {1990})}\BibitemShut {NoStop}%
\bibitem [{\citenamefont {{Schwartz}}\ and\ \citenamefont {{Le
  Roy}}(1987)}]{Schwartz1987}%
  \BibitemOpen
  \bibfield  {author} {\bibinfo {author} {\bibfnamefont {C.}~\bibnamefont
  {{Schwartz}}}\ and\ \bibinfo {author} {\bibfnamefont {R.~J.}\ \bibnamefont
  {{Le Roy}}},\ }\href@noop {} {\bibfield  {journal} {\bibinfo  {journal} {J.
  Mol. Spectrosc.}\ }\textbf {\bibinfo {volume} {121}},\ \bibinfo {pages} {420}
  (\bibinfo {year} {1987})}\BibitemShut {NoStop}%
\bibitem [{\citenamefont {Stanke}\ and\ \citenamefont
  {Adamowicz}(2014)}]{Stanke2014}%
  \BibitemOpen
  \bibfield  {author} {\bibinfo {author} {\bibfnamefont {M.}~\bibnamefont
  {Stanke}}\ and\ \bibinfo {author} {\bibfnamefont {L.}~\bibnamefont
  {Adamowicz}},\ }\href@noop {} {\bibfield  {journal} {\bibinfo  {journal} {J.
  Chem. Phys.}\ }\textbf {\bibinfo {volume} {141}},\ \bibinfo {pages} {154302}
  (\bibinfo {year} {2014})}\BibitemShut {NoStop}%
\bibitem [{\citenamefont {Wolniewicz}(1993)}]{Wolniewicz1993}%
  \BibitemOpen
  \bibfield  {author} {\bibinfo {author} {\bibfnamefont {L.}~\bibnamefont
  {Wolniewicz}},\ }\href@noop {} {\bibfield  {journal} {\bibinfo  {journal}
  {\jcp}\ }\textbf {\bibinfo {volume} {99}},\ \bibinfo {pages} {1851} (\bibinfo
  {year} {1993})}\BibitemShut {NoStop}%
\bibitem [{\citenamefont {Puchalski}\ \emph {et~al.}(2016)\citenamefont
  {Puchalski}, \citenamefont {Komasa}, \citenamefont {Czachorowski},\ and\
  \citenamefont {Pachucki}}]{Puchalski2016}%
  \BibitemOpen
  \bibfield  {author} {\bibinfo {author} {\bibfnamefont {M.}~\bibnamefont
  {Puchalski}}, \bibinfo {author} {\bibfnamefont {J.}~\bibnamefont {Komasa}},
  \bibinfo {author} {\bibfnamefont {P.}~\bibnamefont {Czachorowski}}, \ and\
  \bibinfo {author} {\bibfnamefont {K.}~\bibnamefont {Pachucki}},\ }\href@noop
  {} {\bibfield  {journal} {\bibinfo  {journal} {Phys. Rev. Lett.}\ }\textbf
  {\bibinfo {volume} {117}},\ \bibinfo {pages} {263002} (\bibinfo {year}
  {2016})}\BibitemShut {NoStop}%
\bibitem [{\citenamefont {Niu}\ \emph {et~al.}(2014)\citenamefont {Niu},
  \citenamefont {Salumbides}, \citenamefont {Dickenson}, \citenamefont
  {Eikema},\ and\ \citenamefont {Ubachs}}]{Niu2014}%
  \BibitemOpen
  \bibfield  {author} {\bibinfo {author} {\bibfnamefont {M.~L.}\ \bibnamefont
  {Niu}}, \bibinfo {author} {\bibfnamefont {E.~J.}\ \bibnamefont {Salumbides}},
  \bibinfo {author} {\bibfnamefont {G.~D.}\ \bibnamefont {Dickenson}}, \bibinfo
  {author} {\bibfnamefont {K.~S.~E.}\ \bibnamefont {Eikema}}, \ and\ \bibinfo
  {author} {\bibfnamefont {W.}~\bibnamefont {Ubachs}},\ }\href@noop {}
  {\bibfield  {journal} {\bibinfo  {journal} {\jms}\ }\textbf {\bibinfo
  {volume} {300}},\ \bibinfo {pages} {44} (\bibinfo {year} {2014})}\BibitemShut
  {NoStop}%
\bibitem [{\citenamefont {Angeli}\ and\ \citenamefont
  {Marinova}(2013)}]{Angeli2013}%
  \BibitemOpen
  \bibfield  {author} {\bibinfo {author} {\bibfnamefont {I.}~\bibnamefont
  {Angeli}}\ and\ \bibinfo {author} {\bibfnamefont {K.}~\bibnamefont
  {Marinova}},\ }\href {\doibase https://doi.org/10.1016/j.adt.2011.12.006}
  {\bibfield  {journal} {\bibinfo  {journal} {Atom Data Nucl Data}\ }\textbf
  {\bibinfo {volume} {99}},\ \bibinfo {pages} {69 } (\bibinfo {year}
  {2013})}\BibitemShut {NoStop}%
\bibitem [{\citenamefont {Pohl}\ \emph {et~al.}(2010)\citenamefont {Pohl},
  \citenamefont {Antognini}, \citenamefont {Nez}, \citenamefont {Amaro},
  \citenamefont {Biraben}, \citenamefont {Cardoso}, \citenamefont {Covita},
  \citenamefont {Dax}, \citenamefont {Dhawan}, \citenamefont {Fernandes},
  \citenamefont {Giesen}, \citenamefont {Graf}, \citenamefont {H{\"{a}}nsch},
  \citenamefont {Indelicato}, \citenamefont {Julien}, \citenamefont {Kao},
  \citenamefont {Knowles}, \citenamefont {Bigot}, \citenamefont {Liu},
  \citenamefont {Lopes}, \citenamefont {Ludhova}, \citenamefont {Monteiro},
  \citenamefont {Mulhauser}, \citenamefont {Nebel}, \citenamefont {Rabinowitz},
  \citenamefont {dos Santos}, \citenamefont {Schaller}, \citenamefont
  {Schuhmann}, \citenamefont {Schwob}, \citenamefont {Taqqu}, \citenamefont
  {Veloso},\ and\ \citenamefont {Kottmann}}]{Pohl2010}%
  \BibitemOpen
  \bibfield  {author} {\bibinfo {author} {\bibfnamefont {R.}~\bibnamefont
  {Pohl}}, \bibinfo {author} {\bibfnamefont {A.}~\bibnamefont {Antognini}},
  \bibinfo {author} {\bibfnamefont {F.}~\bibnamefont {Nez}}, \bibinfo {author}
  {\bibfnamefont {F.~D.}\ \bibnamefont {Amaro}}, \bibinfo {author}
  {\bibfnamefont {F.}~\bibnamefont {Biraben}}, \bibinfo {author} {\bibfnamefont
  {J.~M.~R.}\ \bibnamefont {Cardoso}}, \bibinfo {author} {\bibfnamefont
  {D.~S.}\ \bibnamefont {Covita}}, \bibinfo {author} {\bibfnamefont
  {A.}~\bibnamefont {Dax}}, \bibinfo {author} {\bibfnamefont {S.}~\bibnamefont
  {Dhawan}}, \bibinfo {author} {\bibfnamefont {L.~M.~P.}\ \bibnamefont
  {Fernandes}}, \bibinfo {author} {\bibfnamefont {A.}~\bibnamefont {Giesen}},
  \bibinfo {author} {\bibfnamefont {T.}~\bibnamefont {Graf}}, \bibinfo {author}
  {\bibfnamefont {T.~W.}\ \bibnamefont {H{\"{a}}nsch}}, \bibinfo {author}
  {\bibfnamefont {P.}~\bibnamefont {Indelicato}}, \bibinfo {author}
  {\bibfnamefont {L.}~\bibnamefont {Julien}}, \bibinfo {author} {\bibfnamefont
  {C.~Y.}\ \bibnamefont {Kao}}, \bibinfo {author} {\bibfnamefont
  {P.}~\bibnamefont {Knowles}}, \bibinfo {author} {\bibfnamefont {E.-O.~L.}\
  \bibnamefont {Bigot}}, \bibinfo {author} {\bibfnamefont {Y.-W.}\ \bibnamefont
  {Liu}}, \bibinfo {author} {\bibfnamefont {J.~A.~M.}\ \bibnamefont {Lopes}},
  \bibinfo {author} {\bibfnamefont {L.}~\bibnamefont {Ludhova}}, \bibinfo
  {author} {\bibfnamefont {C.~M.~B.}\ \bibnamefont {Monteiro}}, \bibinfo
  {author} {\bibfnamefont {F.}~\bibnamefont {Mulhauser}}, \bibinfo {author}
  {\bibfnamefont {T.}~\bibnamefont {Nebel}}, \bibinfo {author} {\bibfnamefont
  {P.}~\bibnamefont {Rabinowitz}}, \bibinfo {author} {\bibfnamefont {J.~M.~F.}\
  \bibnamefont {dos Santos}}, \bibinfo {author} {\bibfnamefont {L.~A.}\
  \bibnamefont {Schaller}}, \bibinfo {author} {\bibfnamefont {K.}~\bibnamefont
  {Schuhmann}}, \bibinfo {author} {\bibfnamefont {C.}~\bibnamefont {Schwob}},
  \bibinfo {author} {\bibfnamefont {D.}~\bibnamefont {Taqqu}}, \bibinfo
  {author} {\bibfnamefont {J.~F. C.~A.}\ \bibnamefont {Veloso}}, \ and\
  \bibinfo {author} {\bibfnamefont {F.}~\bibnamefont {Kottmann}},\ }\href@noop
  {} {\bibfield  {journal} {\bibinfo  {journal} {Nature}\ }\textbf {\bibinfo
  {volume} {466}},\ \bibinfo {pages} {213} (\bibinfo {year}
  {2010})}\BibitemShut {NoStop}%
\bibitem [{\citenamefont {Antognini}\ \emph {et~al.}(2013)\citenamefont
  {Antognini}, \citenamefont {Nez}, \citenamefont {Schuhmann}, \citenamefont
  {Amaro}, \citenamefont {Biraben}, \citenamefont {Cardoso}, \citenamefont
  {Covita}, \citenamefont {Dax}, \citenamefont {Dhawan}, \citenamefont
  {Diepold}, \citenamefont {Fernandes}, \citenamefont {Giesen}, \citenamefont
  {Gouvea}, \citenamefont {Graf}, \citenamefont {H{\"a}nsch}, \citenamefont
  {Indelicato}, \citenamefont {Julien}, \citenamefont {Kao}, \citenamefont
  {Knowles}, \citenamefont {Kottmann}, \citenamefont {Le~Bigot}, \citenamefont
  {Liu}, \citenamefont {Lopes}, \citenamefont {Ludhova}, \citenamefont
  {Monteiro}, \citenamefont {Mulhauser}, \citenamefont {Nebel}, \citenamefont
  {Rabinowitz}, \citenamefont {dos Santos}, \citenamefont {Schaller},
  \citenamefont {Schwob}, \citenamefont {Taqqu}, \citenamefont {Veloso},
  \citenamefont {Vogelsang},\ and\ \citenamefont {Pohl}}]{Antognini2013}%
  \BibitemOpen
  \bibfield  {author} {\bibinfo {author} {\bibfnamefont {A.}~\bibnamefont
  {Antognini}}, \bibinfo {author} {\bibfnamefont {F.}~\bibnamefont {Nez}},
  \bibinfo {author} {\bibfnamefont {K.}~\bibnamefont {Schuhmann}}, \bibinfo
  {author} {\bibfnamefont {F.~D.}\ \bibnamefont {Amaro}}, \bibinfo {author}
  {\bibfnamefont {F.}~\bibnamefont {Biraben}}, \bibinfo {author} {\bibfnamefont
  {J.~M.~R.}\ \bibnamefont {Cardoso}}, \bibinfo {author} {\bibfnamefont
  {D.~S.}\ \bibnamefont {Covita}}, \bibinfo {author} {\bibfnamefont
  {A.}~\bibnamefont {Dax}}, \bibinfo {author} {\bibfnamefont {S.}~\bibnamefont
  {Dhawan}}, \bibinfo {author} {\bibfnamefont {M.}~\bibnamefont {Diepold}},
  \bibinfo {author} {\bibfnamefont {L.~M.~P.}\ \bibnamefont {Fernandes}},
  \bibinfo {author} {\bibfnamefont {A.}~\bibnamefont {Giesen}}, \bibinfo
  {author} {\bibfnamefont {A.~L.}\ \bibnamefont {Gouvea}}, \bibinfo {author}
  {\bibfnamefont {T.}~\bibnamefont {Graf}}, \bibinfo {author} {\bibfnamefont
  {T.~W.}\ \bibnamefont {H{\"a}nsch}}, \bibinfo {author} {\bibfnamefont
  {P.}~\bibnamefont {Indelicato}}, \bibinfo {author} {\bibfnamefont
  {L.}~\bibnamefont {Julien}}, \bibinfo {author} {\bibfnamefont {C.-Y.}\
  \bibnamefont {Kao}}, \bibinfo {author} {\bibfnamefont {P.}~\bibnamefont
  {Knowles}}, \bibinfo {author} {\bibfnamefont {F.}~\bibnamefont {Kottmann}},
  \bibinfo {author} {\bibfnamefont {E.-O.}\ \bibnamefont {Le~Bigot}}, \bibinfo
  {author} {\bibfnamefont {Y.-W.}\ \bibnamefont {Liu}}, \bibinfo {author}
  {\bibfnamefont {J.~A.~M.}\ \bibnamefont {Lopes}}, \bibinfo {author}
  {\bibfnamefont {L.}~\bibnamefont {Ludhova}}, \bibinfo {author} {\bibfnamefont
  {C.~M.~B.}\ \bibnamefont {Monteiro}}, \bibinfo {author} {\bibfnamefont
  {F.}~\bibnamefont {Mulhauser}}, \bibinfo {author} {\bibfnamefont
  {T.}~\bibnamefont {Nebel}}, \bibinfo {author} {\bibfnamefont
  {P.}~\bibnamefont {Rabinowitz}}, \bibinfo {author} {\bibfnamefont {J.~M.~F.}\
  \bibnamefont {dos Santos}}, \bibinfo {author} {\bibfnamefont {L.~A.}\
  \bibnamefont {Schaller}}, \bibinfo {author} {\bibfnamefont {C.}~\bibnamefont
  {Schwob}}, \bibinfo {author} {\bibfnamefont {D.}~\bibnamefont {Taqqu}},
  \bibinfo {author} {\bibfnamefont {J.~F. C.~A.}\ \bibnamefont {Veloso}},
  \bibinfo {author} {\bibfnamefont {J.}~\bibnamefont {Vogelsang}}, \ and\
  \bibinfo {author} {\bibfnamefont {R.}~\bibnamefont {Pohl}},\ }\href@noop {}
  {\bibfield  {journal} {\bibinfo  {journal} {Science}\ }\textbf {\bibinfo
  {volume} {339}},\ \bibinfo {pages} {417} (\bibinfo {year}
  {2013})}\BibitemShut {NoStop}%
\bibitem [{\citenamefont {Pohl}\ \emph {et~al.}(2016)\citenamefont {Pohl},
  \citenamefont {Nez}, \citenamefont {Fernandes}, \citenamefont {Amaro},
  \citenamefont {Biraben}, \citenamefont {Cardoso}, \citenamefont {Covita},
  \citenamefont {Dax}, \citenamefont {Dhawan}, \citenamefont {Diepold},
  \citenamefont {Giesen}, \citenamefont {Gouvea}, \citenamefont {Graf},
  \citenamefont {H{\"a}nsch}, \citenamefont {Indelicato}, \citenamefont
  {Julien}, \citenamefont {Knowles}, \citenamefont {Kottmann}, \citenamefont
  {Le~Bigot}, \citenamefont {Liu}, \citenamefont {Lopes}, \citenamefont
  {Ludhova}, \citenamefont {Monteiro}, \citenamefont {Mulhauser}, \citenamefont
  {Nebel}, \citenamefont {Rabinowitz}, \citenamefont {dos Santos},
  \citenamefont {Schaller}, \citenamefont {Schuhmann}, \citenamefont {Schwob},
  \citenamefont {Taqqu}, \citenamefont {Veloso},\ and\ \citenamefont
  {Antognini}}]{Pohl2016}%
  \BibitemOpen
  \bibfield  {author} {\bibinfo {author} {\bibfnamefont {R.}~\bibnamefont
  {Pohl}}, \bibinfo {author} {\bibfnamefont {F.}~\bibnamefont {Nez}}, \bibinfo
  {author} {\bibfnamefont {L.~M.~P.}\ \bibnamefont {Fernandes}}, \bibinfo
  {author} {\bibfnamefont {F.~D.}\ \bibnamefont {Amaro}}, \bibinfo {author}
  {\bibfnamefont {F.}~\bibnamefont {Biraben}}, \bibinfo {author} {\bibfnamefont
  {J.~M.~R.}\ \bibnamefont {Cardoso}}, \bibinfo {author} {\bibfnamefont
  {D.~S.}\ \bibnamefont {Covita}}, \bibinfo {author} {\bibfnamefont
  {A.}~\bibnamefont {Dax}}, \bibinfo {author} {\bibfnamefont {S.}~\bibnamefont
  {Dhawan}}, \bibinfo {author} {\bibfnamefont {M.}~\bibnamefont {Diepold}},
  \bibinfo {author} {\bibfnamefont {A.}~\bibnamefont {Giesen}}, \bibinfo
  {author} {\bibfnamefont {A.~L.}\ \bibnamefont {Gouvea}}, \bibinfo {author}
  {\bibfnamefont {T.}~\bibnamefont {Graf}}, \bibinfo {author} {\bibfnamefont
  {T.~W.}\ \bibnamefont {H{\"a}nsch}}, \bibinfo {author} {\bibfnamefont
  {P.}~\bibnamefont {Indelicato}}, \bibinfo {author} {\bibfnamefont
  {L.}~\bibnamefont {Julien}}, \bibinfo {author} {\bibfnamefont
  {P.}~\bibnamefont {Knowles}}, \bibinfo {author} {\bibfnamefont
  {F.}~\bibnamefont {Kottmann}}, \bibinfo {author} {\bibfnamefont {E.-O.}\
  \bibnamefont {Le~Bigot}}, \bibinfo {author} {\bibfnamefont {Y.-W.}\
  \bibnamefont {Liu}}, \bibinfo {author} {\bibfnamefont {J.~A.~M.}\
  \bibnamefont {Lopes}}, \bibinfo {author} {\bibfnamefont {L.}~\bibnamefont
  {Ludhova}}, \bibinfo {author} {\bibfnamefont {C.~M.~B.}\ \bibnamefont
  {Monteiro}}, \bibinfo {author} {\bibfnamefont {F.}~\bibnamefont {Mulhauser}},
  \bibinfo {author} {\bibfnamefont {T.}~\bibnamefont {Nebel}}, \bibinfo
  {author} {\bibfnamefont {P.}~\bibnamefont {Rabinowitz}}, \bibinfo {author}
  {\bibfnamefont {J.~M.~F.}\ \bibnamefont {dos Santos}}, \bibinfo {author}
  {\bibfnamefont {L.~A.}\ \bibnamefont {Schaller}}, \bibinfo {author}
  {\bibfnamefont {K.}~\bibnamefont {Schuhmann}}, \bibinfo {author}
  {\bibfnamefont {C.}~\bibnamefont {Schwob}}, \bibinfo {author} {\bibfnamefont
  {D.}~\bibnamefont {Taqqu}}, \bibinfo {author} {\bibfnamefont {J.~F. C.~A.}\
  \bibnamefont {Veloso}}, \ and\ \bibinfo {author} {\bibfnamefont
  {A.}~\bibnamefont {Antognini}},\ }\href@noop {} {\bibfield  {journal}
  {\bibinfo  {journal} {Science}\ }\textbf {\bibinfo {volume} {353}},\ \bibinfo
  {pages} {669} (\bibinfo {year} {2016})}\BibitemShut {NoStop}%
\end{thebibliography}

%

\end{document}